\documentclass[conference,a4paper]{IEEEtran}

\usepackage{amsmath,amssymb}
\usepackage{bm}
\usepackage{graphicx}
\usepackage{color}

\graphicspath{{figure/}}

\newcommand{\relmiddle}[1]{\mathrel{}\middle#1\mathrel{}} % 伸びる棒
\newcommand{\Prob}[1]{\Pr\left(#1\right)}                 % 確率
\newcommand{\iProb}[1]{\Pr(#1)}                           % 確率（インライン）
\newcommand{\Ex}[1]{{\sf E}\left[#1\right]}               % 期待値
\newcommand{\iEx}[1]{{\sf E}[ #1 ]}                       % 期待値（インライン）
\newcommand{\indicator}[1]{\mathbb{I}\left[#1\right]}     % インジケータ関数
\newcommand{\defeq}{\stackrel{\triangle}{=}}              % 三角イコール
\newcommand{\st}{s{\mathchar`-}t}                         % s-t
\newcommand{\coef}[2]{{\sf coef}\left(#1,#2\right)}       % coef
\newcommand{\icoef}[2]{{\sf coef}(#1,#2)}                 % coef（インライン）
\newcommand{\degree}{d}
\newcommand{\Bgst}{B^{(s,t)}_{G}}
\newcommand{\Agst}{A^{(s,t)}_{G}}
\newcommand{\lgst}{\lambda^{(s,t)}_{G}}
\newcommand{\Cgst}{C^{(s,t)}_{G}}
\newcommand{\cardinality}[1]{\lvert #1 \rvert}
\newcommand{\allgraphs}{R_{n,\degree}^q}
\newcommand{\ensemble}{(\allgraphs, P)}
\newcommand{\cutv}[1]{cut(#1)} % カットベクトル
\newcommand{\cutsetv}[1]{cutset(#1)} % カットセットベクトル

\newtheorem{lemma}{Lemma}
\newtheorem{theorem}{Theorem}

\begin{document}

\sloppy

%% Paper Title
%% You can use linebreaks \\ within to get better formatting as
%% desired.
\title{An Analysis on Minimum $\st$ Cut Capacity of Random Graphs with
Specified Degree Distribution}

%% Author names and affiliations:
%%
%% Avoiding spaces at the end of the author lines is not a problem with
%% conference papers because we don't use \thanks or \IEEEmembership.
%%
%% For several authors with only one affiliation:
%%
% \author{
%   \IEEEauthorblockN{Hui-Ting Chang and Stefan M.~Moser}
%   \IEEEauthorblockA{Department of Electrical and Computer Engineering\\
%     National Chiao Tung University (NCTU)\\
%     Hsinchu, Taiwan\\
%     Email: \{email-of-hui-ting,email-of-stefan\}@ieee.org}
% }
%%
%% For up to three affiliations:
%%
\author{ \IEEEauthorblockN{Yuki Fujii and Tadashi Wadayama}
  \IEEEauthorblockA{Department of Computer Science and Engineering, \\
  Nagoya Institute of Technology, Nagoya, Japan\\ Email: fujii@it.cs.nitech.ac.jp,
  wadayama@nitech.ac.jp} }
  %%
  %% For over three affiliations, or if they all won't fit within the width
  %% of the page, use this alternative format:
  %%
  % \author{
  %   \IEEEauthorblockN{
  %     Michael Shell\IEEEauthorrefmark{1},
  %     Homer Simpson\IEEEauthorrefmark{2},
  %     James Kirk\IEEEauthorrefmark{3},
  %     Montgomery Scott\IEEEauthorrefmark{3} and
  %     Eldon Tyrell\IEEEauthorrefmark{4}}
  %   \IEEEauthorblockA{
  %     \IEEEauthorrefmark{1}School of Electrical and Computer Engineering\\
  %     Georgia Institute of Technology, Atlanta, Georgia 30332--0250\\
  %     Email: see http://www.michaelshell.org/contact.html}
  %   \IEEEauthorblockA{
  %     \IEEEauthorrefmark{2}Twentieth Century Fox, Springfield, USA\\
  %     Email: homer@thesimpsons.com}
  %   \IEEEauthorblockA{
  %     \IEEEauthorrefmark{3}Starfleet Academy, San Francisco, California 96678-2391\\
  %     Telephone: (800) 555--1212, Fax: (888) 555--1212}
  %   \IEEEauthorblockA{
  %     \IEEEauthorrefmark{4}Tyrell Inc., 123 Replicant Street, Los Angeles, California 90210--4321}
  % }

  %% Use for special paper notices
  %\IEEEspecialpapernotice{(Invited Paper)}

  %% To balance the two columns, you should reduce the text-height of
  %% the last page using the following command:
  %%%%%%%%%%%%%%%%%%%%%%%%%%%%%%%%%%%%%%%%%%%%%%%%%%%%%%%%%%%%%%%%%%%%%
  %\addtolength{\textheight}{-9.35cm}
  %%%%%%%%%%%%%%%%%%%%%%%%%%%%%%%%%%%%%%%%%%%%%%%%%%%%%%%%%%%%%%%%%%%%%
  %% with an appropriate value. This command must be place on the second
  %% last page, i.e., for a one-page abstract here, for a two-page
  %% abstract right after the \maketitle command.

  %% Create the title:
  \maketitle

  %% Abstract:
  %% For the final version of the accepted paper, please make sure you
  %% remove the comment "THIS PAPER IS ELIGIBLE FOR THE STUDENT PAPER
  %% AWARD."
  %%
  \begin{abstract}
   The capacity (or maximum flow) of an unicast network is known to be
   equal to the minimum $\st$ cut capacity due to the max-flow min-cut
   theorem. If the topology of a network (or link capacities) is
   dynamically changing or unknown, it is not so trivial to predict
   statistical properties on the maximum flow of the network. In this
   paper, we present a probabilistic analysis for evaluating the
   accumulate distribution of the minimum $\st$ cut capacity on random
   graphs. The graph ensemble treated in this paper consists of weighted
   graphs with arbitrary specified degree distribution. The main
   contribution of our work is a lower bound for the accumulate
   distribution of the minimum $\st$ cut capacity. From some computer
   experiments, it is observed that the lower bound derived here
   reflects the actual statistical behavior of the minimum $\st$ cut
   capacity of random graphs with specified degrees.
  \end{abstract}

  \section{Introduction}

  Rapid growth of information flow over a network such as a backbone
  network for mobile terminals requires efficient utilization of full
  potential of the network. In a multicast communication scenario, it is
  well known that appropriate network coding achieves its multicast
  capacity. Emergence of the network coding have
  broaden network design strategies for efficient use of wired and
  wireless networks \cite{Ahlswede2000}.

  The multicast capacity of a directed graph is closely related to the
  $\st$ maximum flow, which is equal to the minimum $\st$ cut capacity
  due to the max-flow min-cut theorem \cite{Schrijver2003}. Furthermore,
  on a unicast network, the minimum $\st$ cut capacity of the network
  determines the unicast capacity between the terminals $s$ and $t$.
  Therefore, it is meaningful to study the minimum $\st$ cut capacity
  for designing an efficient network.

  If the topology of a network is static, the corresponding $\st$
  maximum flow of the network can be efficiently evaluated in polynomial
  time by using Ford-Fulkerson algorithm \cite{Schrijver2003}. However,
  if the topology of a network and its link capacities are dynamically
  changing or have stochastic nature, it is not so trivial to predict
  statistical properties on the maximum flow. For example, in a case of
  wireless network, the link capacities may fluctuate because of the
  effect of time-varying fading. Another example is an ad-hoc network
  whose link connections are stochastically determined.

  In order to obtain an insight for statistical properties of the
  minimum $\st$ cut capacity for such random networks, it is natural to
  investigate statistical properties of minimum $\st$ cut capacity over
  a random graph ensemble.  Such a result may unveil typical behaviors
  of the minimum $\st$ cut capacity (or maximum flow) for given
  parameters of a network such as the number of vertices, edges,
  probabilistic properties of edge weight and degree distributions.

  Several theoretical works on the maximum flow of random graphs (i.e.,
  graph ensembles) have been made. In a context of randomized
  algorithms, Karger showed a sharp concentration result for maximum
  flow in the asymptotic regime \cite{Karger1999}. Ramamoorthy et al.
  presented another concentration result. The network coding capacities
  of weighted random graphs and weighted random geometric graphs
  concentrate around the expected number of nearest neighbors of the
  source and the sinks \cite{Ramamoorthy2005a}. These concentration
  results indicate an asymptotic properties of the maximum flow of
  random networks. Wang et al. shows statistical properties of the
  maximum flow in an asymptotic setting as well. They discussed the
  random graphs with Bernoulli distributed weights \cite{Wang2007a}.

  In this paper, we will present a lower bound for the accumulate
  distribution of the minimum $\st$ cut capacity of weighted random
  graphs with specified degree distribution. The approach presented here
  is totally different from those used in the conventional works
  \cite{Karger1999}\cite{Ramamoorthy2005a}\cite{Wang2007a}. The basis of
  the analysis is the correspondence between the cut space of an
  undirected graph and a binary LDGM (low-density generator-matrix) code
  \cite{Hakimi1965}. Based on this correspondence, Yano and Wadayama
  \cite{Yano-isita2012} presented an ensemble analysis for the network
  reliability problem.  Fujii and Wadayama \cite{Fujii-isita2012}
  proposed a probabilistic analysis for the global minimum cut capacity
  over the weighted Erd\H{o}s-R\'{e}nyi random graphs. The probability
  distribution of vertex degrees over Erd\H{o}s-R\'{e}nyi random graphs
  follows the Poisson distribution. However, most of degree
  distributions of real networks are different from the Poisson
  distribution \cite{barabasi2002linked}.  This paper extends the idea
  in \cite{Yano-isita2012} and \cite{Fujii-isita2012} to weighted random
  graphs with arbitrary specified degree distribution, which may be
  applicable to more realistic networks.  Moreover, this paper deals
  with $\st$ cut capacity which is more informative on network
  capacities instead of the global cut capacity \cite{Fujii-isita2012}.

  \section{Preliminaries}

  In this section, we first introduce several basic definitions and
  notation used throughout the paper. Then, an ensemble of weighted
  undirected graphs treated in this paper is defined.

  \subsection{Notation and definitions}

  A graph $G\defeq(V,E)$ is a pair of a vertex set $V \defeq \{v_1, v_2,
  \ldots, v_n\}$ and an edge set $E \defeq \{e_1, e_2,\ldots, e_m \}$
  where $e_j = (u,v), u,v \in V$ is an edge.  If $e_j = (u,v)$ is not an
  ordered pair, i.e., $(u,v)=(v,u)$, the graph $G$ is called an
  \emph{undirected graph}.

  If a function $c:E \rightarrow \mathbb{Z}_{\ge 0}$ is defined for an
  undirected graph $G\defeq(V,E)$, the triple $(V, E, c)$ is considered
  as a \emph{weighted graph}. The function $c$ can be seen as weight for
  edges. The set $\mathbb{Z}_{\ge 0}$ represents the set of non-negative
  integers. In our context, the weight function $c$ represents the link
  capacity for each edge.

  Assume that a weighted undirected graph $G\defeq(V, E, c)$ is given. A
  non-overlapping bi-partition $V=X \cup (V \backslash X)$ is called a
  \emph{cut} where $X$ is a non-empty proper subset of $V (X \ne
  V)$. The set of edges bridging $X$ and $V \backslash X$ is referred to
  as the \emph{cut-set} corresponding to the cut $(X, V \backslash X)$,
  which is denoted by $\partial(X)$ (or equivalently
  $\partial(V\backslash X)$).  The \emph{cut weight} (i.e., cut
  capacity) of $X$ is defined as $\omega(X) \defeq \sum_{e \in
  \partial(X)} c(e).$ If a cut $(X, V \backslash X)$ separates two
  vertices $s,t \in V (s \neq t)$, the cut $(X, V \backslash X)$ is
  called an \emph{$\st$ cut} and the corresponding cut-set is called an
  \emph{$\st$ cut-set}. The \emph{minimum $\st$ cut} is an $\st$ cut
  whose cut weight is the smallest among all the $\st$ cut-sets.

  \subsection{Random graphs with specified degree distribution}
  \label{section:Random Graphs}

  In the following, we will define an ensemble of weighted undirected
  graphs. The random graph ensemble is a weighted version of random
  graphs with arbitrary specified degree distribution treated in
  \cite{Newman2001}. Let $n$ ($n \ge 1$) be the number of vertices and
  $\degree_i$ be the fraction of vertices having degree $i$ such that $n
  \degree_i$ is an non-negative integer and $\sum_{i=1}^{\infty} i n
  \degree_i$ is even. We define $\degree(x) \defeq \sum_{i = 1}^{\infty}
  \degree_i x^{i}$ to be the generating function of $\degree_i$. Due to
  these assumptions, the number of edges $m$ is given by $1/2
  \sum_{i=1}^{\infty} i n \degree_i$.

  It is assumed that each edge has own integer weight; namely, a weight
  $w_i \in [1,q]$ ($i \in [1,m]$) is assigned to the $i$th edge. The
  notation $[a,b]$ denotes the set of consecutive integers from $a$ to
  $b$. The set $\allgraphs$ denotes the set of all the undirected
  weighted graphs satisfying the above assumption.

  We here assign the probability
  \begin{equation}
   P(G) \defeq \frac{1}{\cardinality{\allgraphs}}
    \prod_{e \in E} \mu(c(e))
  \end{equation}
  for $G \in \allgraphs$ where $\mu$ is a discrete probability measure
  defined over $[1,q]$; namely, it satisfies $\sum_{w\in [1,q] }\mu(w) =
  1 \text{ and } \forall w \in [1,q], \mu(w) \ge 0.$ The pair
  $\ensemble$ defines an ensemble of random graphs treated in this
  paper.

 \section{Cut Weight Distribution}

  \subsection{Constraint graph}

  In this paper, we use a bipartite graph, which is called a
  \emph{constraint graph}\footnote{A constraint graph can be considered
  as a factor graph.}, corresponding to a given undirected graph. The
  constraint graph clarifies the close relationship between the
  incidence vectors of cut and cut-sets.  In the following, we will
  explain the definition of the constraint graph $G' \defeq (V_1, V_2,
  E')$ corresponding to an undirected graph $G \defeq (V,E)$.

\begin{figure}[!t]
   \centering
   \includegraphics[width=1.0\columnwidth]{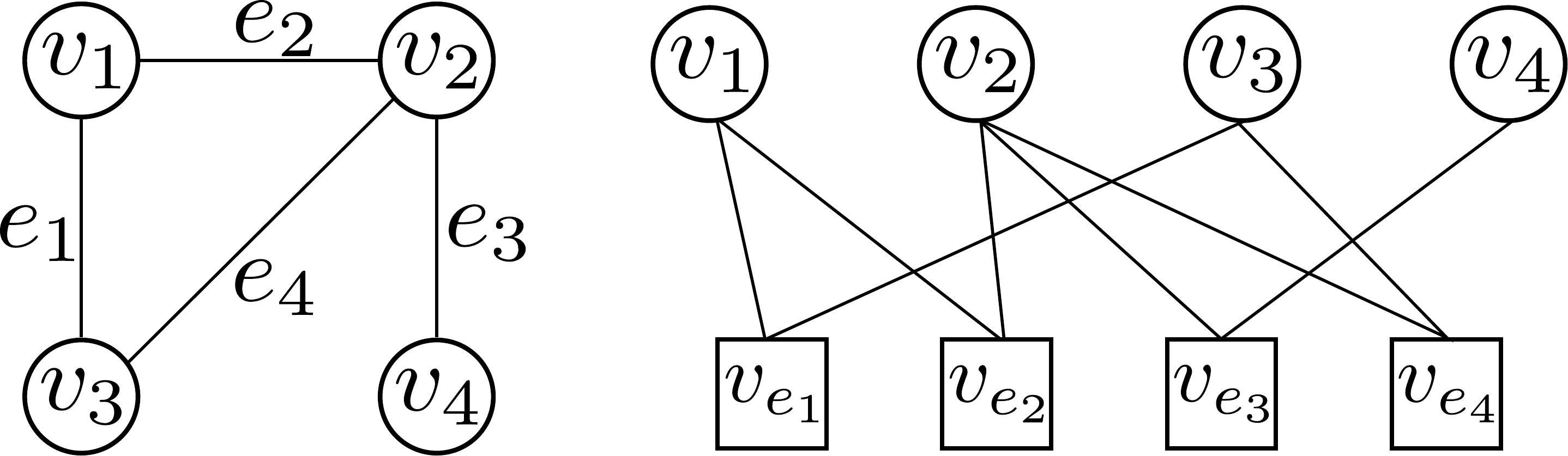}
   \caption{An undirected graph (left) and corresponding constraint graph (right)}
   \label{fig:Example of conversion}
  \end{figure}

  Suppose that an undetected graph $G$ is given. In order to construct
  the constraint graph from $G$, for each edge $e=(x,y) \in E$, we
  insert a new vertex $v_e$ between $x$ and $y$. The new vertex $v_e$
  is, thus, adjacent to $x$ and $y$. Formally, the triple $(V_1, V_2,
  E')$ for the constraint graph $G'$ is defined by
  \begin{align}
   &V_1 \defeq V, \; \; V_2 \defeq \left\{ v_{e_i} \relmiddle| e_i \in E
   \right\},\notag \\
   &E' \defeq \left\{ (x, v_{e_i}) , (y, v_{e_i})
   \relmiddle| e_i = (x,y) \in E \right\}.
  \end{align}
  From this definition, it is clear that the degree of all vertices in
  $V_2$ is $2$. Figure \ref{fig:Example of conversion} illustrates the
  correspondence between the original graph (left) and the constraint
  graph (right).

  \subsection{Relationship between cut-set vector and cut vector}

  For a given undirected graph $G \defeq (V,E)$, the \emph{cut vector}
  $\cutv{X}\defeq(a_1,\ldots, a_n)$ of a cut $(X, V \backslash X)$ is
  defined by $a_i \defeq \indicator{v_i \in X}$ for $i \in [1,n]$. The
  function $\indicator{\cdot}$ is the indicator function that takes
  value 1 if the condition is true; otherwise it takes value 0. Namely,
  the cut vector $\cutv{X}$ is the incidence vector of the cut $(X, V
  \backslash X)$.  In a similar manner, we will define the
  \emph{cut-set} vector as follows.  The cut-set vector $\cutsetv{X}
  \defeq (b_1,\ldots, b_m)$ corresponding to a cut $(X, V \backslash X)$
  is defined by $ b_i \defeq \indicator{e_i \in \partial(X)} $ for $i
  \in [1,m]$.

  The constraint graph naturally connects a cut vector $\cutv{X}$ and
  the corresponding cut-set vector $\cutsetv{X}$ for any $X \subset V (X
  \neq \emptyset)$ in the following way. Suppose that an undirected
  graph $G \defeq (V,E)$ and the corresponding constraint graph $G'
  \defeq (V_1, V_2, E')$ are given. The vertices in $V_1$ are called
  \emph{variable nodes} which are depicted by circles in
  Fig.\ref{fig:Example of conversion}.  We assume that a binary value (0
  or 1) can be assigned to a variable node. The vertices in $V_2$ are
  called \emph{function nodes} which are represented by squares in
  Fig.\ref{fig:Example of conversion}. The function node also have a
  binary value which is determined by the bitwise exclusive-OR (sum over
  $\mathbb{F}_2$) of values in adjacent variable nodes. Let us assume
  that $\bm{x} \defeq (x_1,\ldots, x_n) \in \{0,1\}^n$ is assigned to
  the variable nodes (i.e., $x_i$ is the assigned value for $v_i$) and
  that $\bm{y} \defeq (y_1,\ldots, y_m) \in \{0,1\}^m$ is the resulting
  values (i.e., $y_i$ is the exclusive-OR value at $v_{e_i}$). The
  linear relation between $\bm{x}$ and $\bm{y}$ is denoted by $\bm{y} =
  F_G(\bm{x})$. The next lemma presents the linear relation between a
  cut vector and the corresponding cut-set vector.
  \begin{lemma}
   \label{linear} Assume that an undirected graph $G\defeq(V,E)$ is
   given. For any $X \subset V (X \neq \emptyset)$, the following linear
   relation
   \begin{equation}
    \cutsetv{X} = F_G(\cutv{X})
   \end{equation}
   holds.
  \end{lemma}

  \begin{IEEEproof}
   Let $(y_1,\ldots, y_m) \defeq F_G(\cutv{X})$ be a vector at the
   function nodes and $G' \defeq (V_1, V_2, E')$ be the constraint graph
   corresponding to $G$. Two variable nodes adjacent to $v_{e_i}$ are
   denoted by $a, b \in V_1$.  If $a \in X, b \in V\backslash X$, then
   $y_i = 1$. Otherwise, $y_i = 0$.  From the definition of the
   constraint graph, $y_i =1$ is equivalent to $e_i \in \partial (X)$.
   This proves the relation $\cutsetv{X} = F_G(\cutv{X})$.
  \end{IEEEproof}

  It should be remarked that the linear relation in Lemma \ref{linear}
  has been long known in the field of graph theory; e.g.,
  \cite{Hakimi1965}.  Namely, a linear row space spanned by the
  incidence matrix of $G$ coincides with the set of incidence vectors of
  cut-sets.

  \subsection{$\st$ cut weight distribution}

  Assume that a weight undirected graph $G \defeq (V,E,c)$ and two
  vertices $s,t \in V (s \neq t)$ are given. The $\st$ \emph{cut weight
  distribution} is defined by
  \begin{equation}
   \Bgst(w)
    \defeq \sum_{E' \subseteq E} \indicator{
    \text{$E'$ is an $\st$ cut-set}, \sum_{e \in E'} c(e)=w}
  \end{equation}
  for non-negative integer $w$. The $\st$ cut weight distribution
  $\Bgst(w)$ represents the number of cut-sets with cut weight $w$. The
  following lemma plays an important role for evaluating the ensemble
  average of the cut weight distribution $\Bgst(w)$.
  \begin{lemma}
   \label{bwgst} The $\st$ cut weight distribution $\Bgst(w)$ can be
   upper bounded by
   \begin{equation}
    \Bgst(w) \le \frac{1}{2} \sum_{u=1}^{n-1} \sum_{v=0}^m \Agst(u,v,w),
   \end{equation}
   for $w \in \mathbb{Z}_{\ge 0}$. The quantity $\Agst(u,v,w)$ is
   defined by
   \begin{align}
    &\Agst(u,v,w) \nonumber\\
    &\defeq \!\!\!\!
    \sum_{\bm{a} \in Y^{(s,t)} \cap Z^{(n,u)}} \sum_{\bm{b} \in Z^{(m,v)}}
    \indicator{F_G(\bm{a}) = \bm{b}, \sum_{i=1}^m b_i c(e_i) = w},
   \end{align}
   for $u \in [1,n-1]$, $v \in [0,m]$ and $w \in \mathbb{Z}_{\ge 0}$.
   The set of the constant weight binary vectors $Z^{(x,y)}$ is defined
   as $Z^{(x,y)} \defeq \left\{(z_1, \ldots, z_x) \in \{0,1\}^x \relmiddle|
   \sum_{i=1}^x z_i = y \right\}$. The set $Y^{(s,t)}$ denotes the set
   of all $\st$ cut vectors.
  \end{lemma}

  \begin{IEEEproof}
   For any undirected graph $G\defeq(V,E)$, $\Bgst(w)$ can be upper
   bounded by
   \begin{equation}
    \label{Bwineq}
    \Bgst(w)
    \le \frac{1}{2} \sum_{X \subset V, X \neq \emptyset}
    \indicator{\text{$X$ is an $\st$ cut,\ } \omega(X)=w}.
   \end{equation}
   The factor $1/2$ is required for compensating the double counting for
   $X$ and $V \backslash X$. The equality is attained if and only if $G$
   is connected. Due to Lemma \ref{linear}, the right-hand side of
   (\ref{Bwineq}) can be rewritten as
   \begin{multline}
    \frac{1}{2}\sum_{X \subset V, X \neq \emptyset}
    \indicator{\text{$X$ is an $\st$ cut,\ } \omega(X)=w} \\
    = \frac{1}{2} \sum_{u = 1}^{n-1} \sum_{v = 0}^m \Agst(u,v,w).
    \label{eq:右辺を変形}
   \end{multline}
   Substituting \eqref{eq:右辺を変形} into (\ref{Bwineq}), we obtain the
   claim.
  \end{IEEEproof}

  \section{Ensemble average of $\st$ cut weight distribution}

  In this section, we will discuss the ensemble average of $\Bgst(w)$
  over the ensemble $\ensemble$.

  \subsection{Upper bound on average cut weight distribution}

  Due to the linearity of the expectation over the ensemble and Lemma
  \ref{bwgst}, we have
  \begin{equation}
   \Ex{\Bgst(w)} \le \frac{1}{2}
    \sum_{u = 1}^{n-1} \sum_{v = 0}^{m} \Ex{\Agst(u,v,w)}.
    \label{eq:平均カット重み分布と平均入出力重み分布}
  \end{equation}
  In the following, we will analyze $\iEx{\Agst(u,v,w)}$. The analysis
  presented below is similar to the derivation of the average
  input-output weight distribution of irregular LDGM codes due to Hsu
  and Anastasopoulos \cite{Anastasopoulos2005a}. The next lemma provides
  the expectation of $\Agst(u,v,w)$ by using the generating function
  method.
  \begin{lemma}
   \label{lemma:auvw} For any pair of $s$ and $t$ ($s \neq t$), the
   expectation of $\Agst(u,v,w)$ over $\ensemble$ is given by
   \begin{align}
    &\hspace{-20pt}
    \Ex{\Agst(u,v,w)} \nonumber\\
    =&\frac{
    2^{v+1}u(n-u) \binom{m}{v} \coef{f(x)^v}{x^w}
    }
    {
    n(n-1)
    } \nonumber\\
    &\times \sum_{h=0}^{2m}
    \frac{
    \binom{m-v}{\frac{h-v}{2}}
    \coef{\prod_{i=1}^{\infty} (1+x^i y)^{n \degree_i}}{x^h y^u}
    }{
    \binom{2m}{h}
    },
    \label{eq:auvw}
   \end{align}
   where $u \in [1,n-1]$, $v \in [0,m]$, $w \in \mathbb{Z}_{\ge 0}$. The
   generator function $f(x)$ is defined by $f(x) \defeq \sum_{i=1}^q
   \mu(i) x^i.$ The notation $\icoef{f(x,y)}{x^a y^b}$ represents the
   coefficient of $x^a y^b$ in the polynomial $f(x,y)$.
  \end{lemma}

  \begin{IEEEproof}
   The expectation of $\Agst(u,v,w)$ can be simplified as follows:
   \begin{align}
    &\Ex{\Agst(u,v,w)} \nonumber\\
    =& \!\!\!\!\!\!\!\sum_{\bm{a} \in Y^{(s,t)} \cap Z^{(n,u)}}
    \sum_{\bm{b} \in Z^{(m,v)}}
    \!\!\!\Ex{\indicator{F_G(\bm{a}) = \bm{b}, \sum_{i=1}^m b_i c(e_i) = w}}
    \nonumber \\
    =& 2 \binom{n-2}{u-1} \binom{m}{v}
    \Ex{\indicator{F_G(\bm{a}^*)
    = \bm{b}^*, \sum_{i=1}^m b^*_i c(e_i) = w}}, \label{eq:indicator-ex-a}
   \end{align}
   where binary vectors $\bm{a}^* \in Y^{(s,t)} \cap Z^{(n,u)}$ and
   $\bm{b}^* \in Z^{(m,v)}$. The last equality is due to the symmetry of
   the ensemble. The expectation in \eqref{eq:indicator-ex-a} can be
   rewritten as follows:
   \begin{align}
    &\Ex{\indicator{F_G(\bm{a}^*)
    = \bm{b}^*, \sum_{i=1}^m b^*_i c(e_i) = w}} \nonumber\\
    =& \sum_{G \in \allgraphs} P(G) \;
    \indicator{F_G(\bm{a}^*) = \bm{b}^*, \sum_{i=1}^m b^*_i c(e_i) = w}
    \nonumber \\
    =& \Prob{B = \bm{b}^*, W = w \relmiddle| A = \bm{a}^*} \nonumber\\
    =& \Prob{B = \bm{b}^* \relmiddle| A = \bm{a}^*}
    \Prob{W = w \relmiddle| B = \bm{b}^*, A = \bm{a}^*},
    \label{eq:indicator-ex-b}
   \end{align}
   where $A$, $B$ and $W$ are random variables representing a cut
   vector, a cut-set vector and cut weight, respectively.

   Edges connecting to variable nodes having value $1$ are referred to
   as \emph{active edges}. Let $H$ be the random variable of the total
   number of active edges. Since the number of all edges between
   variable nodes and function nodes is $2m$, we have
   \begin{align}
    &\Prob{B = \bm{b}^* \relmiddle| A = \bm{a}^*} \nonumber\\
    =& \sum_{h = 0}^{2m}
    \Prob{B = \bm{b}^*, H = h \relmiddle| A = \bm{a}^*} \nonumber\\
    =& \sum_{h = 0}^{2m}
    \Prob{B = \bm{b}^* \relmiddle| H = h, A = \bm{a}^*}
    \Prob{H = h \relmiddle| A = \bm{a}^*}.
    \label{eq:indicator-ex-c}
   \end{align}
%   （日本語：カットベクトルが$\bm{m}^*$のとき，値が$1$の変
%   数ノード$u$個に$h$本の辺が接続するパターン数は，$\binom{n}{u}$通りのう
%   ち，$\coef{\prod_{i=1}^{\infty} (1+x^i y)^{n \degree_i}}{x^h
%   y^u}$通りと計算できるので，）
   Since the number of ways that the cut vector is $\bm{a}^*$ and $h$
   edges connect to $u$ variable nodes having active value, out of a
   total of $\binom{n}{u}$ possibilities, is equal to $
   \icoef{\prod_{i=1}^{\infty} (1+x^i y)^{n \degree_i}}{x^h y^u}, $ we
   have
   \begin{equation}
    \Prob{H = h \relmiddle| A = \bm{a}^*}
    = \frac{
    \coef{\prod_{i=1}^{\infty} (1+x^i y)^{n\degree_i}}{x^h y^u}
    }{
    \binom{n}{u}
    }.
    \label{eq:indicator-ex-d}
   \end{equation}
%   また，カットセットベクトルが$\bm{c}^*$になるためには，$v$個の値1の関数
%   ノードに1本のアクティブエッジが接続し，残りの$m-v$個の関数ノードには，
%   2本のアクティブエッジが接続するか，アクティブエッジが接続してはいけな
%   い．アクティブエッジが$h$本のとき，この条件を満たすパターン数は，
%   $\binom{2m}{h}$のうち，$2^v \binom{m-v}{(h-v)/2}$通りである．したがっ
%   て，
   A function node with the value $1$ is connected to only one active
   edge because the value of a function node is given by exclusive-OR of
   values of the adjacent variable nodes.  Since the weight of the
   cut-set vector $\bm{b}^*$ is $v$, the number of such function nodes
   with the value $1$ is $v$ and remaining $m-v$ function nodes have the
   value 0. Note that a function node with the value $0$ is connected to
   two active edges or to no active edges.  When the number of all
   active edges is $h$, the number of ways satisfying the above
   condition, out of a total of $\binom{2m}{h}$, is $2^v
   \binom{m-v}{(h-v)/2}$. Therefore, we have
   \begin{equation}
    \Prob{B = \bm{b}^* \relmiddle| H = h, A = \bm{a}^*} =
     \frac{2^v \binom{m-v}{\frac{h-v}{2}}}{\binom{2m}{h}}.
     \label{eq:indicator-ex-f}
   \end{equation}
   Note that this probability is independent of  the cut vector $\bm{a}^*$.

   Since the probability which the cut weight is $w$ depends only on the
   cardinality of the cut-set, we have
   \begin{align}
    &\Prob{W = w \relmiddle| B = \bm{b}^*, A = \bm{a}^*} \notag\\
    =&
    \sum_{
    \substack{p_1 + p_2 + \cdots + p_q = v \\ p_1 + 2 p_2 +\cdots + q p_q = w}
    }
    \binom{v}{p_1, p_2, \ldots, p_q}
    \prod_{i \in [1,q]} \mu(i)^{p_i}
    \notag\\
    =& \coef{f(x)^v}{x^w}.
    \label{eq:indicator-ex-e}
   \end{align}
   The last equality is due to the multinomial theorem. Combining
   \eqref{eq:indicator-ex-a}, \eqref{eq:indicator-ex-b},
   \eqref{eq:indicator-ex-c},
   \eqref{eq:indicator-ex-d},\eqref{eq:indicator-ex-f} and
   \eqref{eq:indicator-ex-e}, we obtain the lemma.
  \end{IEEEproof}

  As a special case of Lemma \ref{lemma:auvw}, if $\degree(x) = x^{c}$
  (i.e., $G$ is a $c$-regular graph), we have
  \begin{equation}
   \Ex{\Agst(u,v,w)} =
    \frac{
    2^{v+1} \binom{n-2}{u-1} \binom{m}{v} \binom{m-v}{\frac{cu-v}{2}}
    \coef{f(x)^v}{x^w}
    }{
    \binom{cn}{cu}
    }.
  \end{equation}

  In order to investigate statistical properties of the minimum $\st$
  cut weight, it is natural to study the tail of the average $\st$ cut
  weight distribution. The following theorem provides an upper bound on
  average cut weight distribution that is the basis of our analysis.

  \begin{theorem}
   \label{thorem:bw} For any pair of $s$ and $t$ ($s \neq t$), the
   expectation of $\Bgst(w)$ over $\ensemble$ can be upper bounded by
   \begin{align}
    \Ex{\Bgst(w)} \le& \sum_{u = 1}^{n-1} \sum_{v = 0}^{m}
    \frac{2^v u(n-u) \binom{m}{v} \coef{f(x)^v}{x^w}}
    {n(n-1)} \notag\\
    &\times \! \sum_{h=0}^{2m}
    \frac{
    \binom{m-v}{\frac{h-v}{2}}
    \coef{\prod_{i=1}^{\infty} (1+x^i y)^{n \degree_i}}{x^h y^u}
    }{
    \binom{2m}{h}
    }.
    \label{eq:theorem:bw}
   \end{align}
  \end{theorem}

  \begin{IEEEproof}
   Applying Lemma \ref{lemma:auvw} to the inequality \eqref{eq:平均カット%
   重み分布と平均入出力重み分布}, we obtain the claim of this theorem.
  \end{IEEEproof}

  \subsection{Minimum $\st$ cut weight}

  Let $\lgst$ be the minimum $\st$ cut weight of the graph $G$ and
  $\Cgst(\delta) \defeq \sum_{w=0}^{\delta-1} \Bgst(w)$ be the
  accumulate $\st$ cut weight of $G$ where $\delta$ is a positive
  integer. From this definition, it is clear that the graph $G$ does not
  contain an $\st$ cut with weight smaller than $\delta$ if
  $\Cgst(\delta)$ is zero. This implies that $\Cgst(\delta)=0$ is
  equivalent to $\lgst \ge \delta$ and that
\[
   \iProb{\lgst \ge \delta} =
    \iProb{\Cgst(\delta) = 0}
    = 1 - \iProb{\Cgst(\delta) \ge 1}.
\]
%  \begin{align}
%   \iProb{\lgst \ge \delta} =& \iProb{\Cgst(\delta) = 0} \notag\\
%   =& 1 - \iProb{\Cgst(\delta) \ge 1}.
%  \end{align}
  The second equality is due to the non-negativity of $\Cgst(\delta)$.

  The following theorem is the main contribution of this work.
  \begin{theorem}
   \label{theorem:maintheorem} The probability $\iProb{\lgst \ge
   \delta}$ can be lower bounded by
   \begin{align}
    &\hspace{-25pt}\iProb{\lgst \ge \delta} \notag\\
    \ge& 1 - \sum_{w = 0}^{\delta - 1}
    \sum_{u = 1}^{n-1} \sum_{v = 0}^{m}
    \frac{2^v u(n-u) \binom{m}{v} \coef{f(x)^v}{x^w}}
    {n(n-1)} \notag\\
    &\times \sum_{h=0}^{2m}
    \frac{
    \binom{m-v}{\frac{h-v}{2}}
    \coef{\prod_{i=1}^{\infty} (1+x^i y)^{n\degree_i}}{x^h y^u}
    }{
    \binom{2m}{h}
    }
   \end{align}
   for $\delta \in \mathbb{N}$ over the ensemble $\ensemble$. The set
   $\mathbb{N}$ represents the set of positive integers.
  \end{theorem}

  \begin{IEEEproof}
   The Markov inequality provides an lower bound on $\iProb{\lgst \ge
   \delta}$ as follows:
   \begin{align}
    \iProb{\lgst \ge \delta} = & 1- \iProb{\Cgst(\delta) \ge 1} \notag \\
    \ge& 1-  \iEx{\Cgst(\delta)}  = 1 - \sum_{w=0}^{\delta-1} \iEx{\Bgst(w)}.
    \label{eq:maintheorem}
   \end{align}
   Applying the lower bound \eqref{eq:theorem:bw} in Theorem
   \ref{thorem:bw} to the inequality \eqref{eq:maintheorem}, we obtain
   the claim of this theorem.
  \end{IEEEproof}

  \section{Numerical result}

  In order to evaluate the tightness of the lower bound shown in Theorem
  \ref{theorem:maintheorem}, we made the following computer
  experiments. In an experiment, we generated $10^4$-instances of
  undirected graphs from the random graph ensemble defined in the
  Section \ref{section:Random Graphs}. We assumed that the edge weight
  is $1$; namely, $q=1$, $\mu(1)=1$. The minimum $\st$ cut weight for
  each instance was computed by using the Ford-Fulkerson algorithm
  \cite{Schrijver2003}.

\begin{figure}[!t]
   \centering
   \includegraphics[width=1.0\columnwidth]{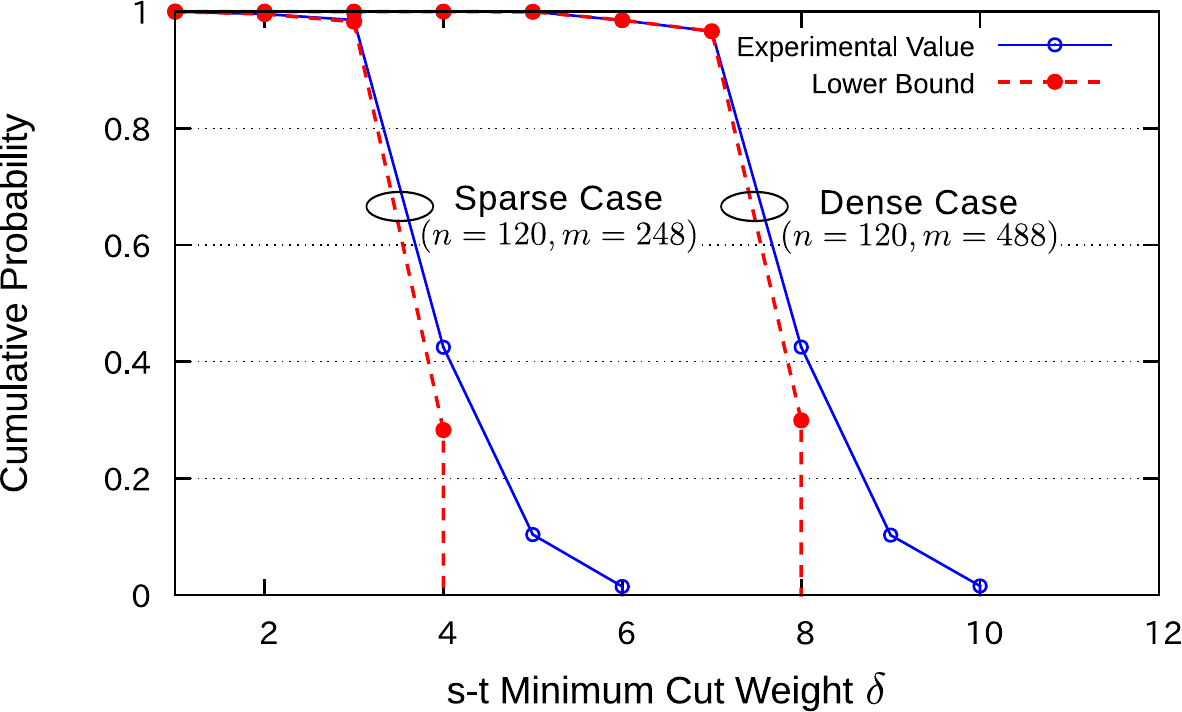}
   \caption{Accumulate distribution of the minimum $\st$ cut weight
   $\iProb{\lgst \ge \delta}$ (sparse case and dense case):
   experimental values and lower bounds.}
   \label{fig:experimental1}
  \end{figure}

\begin{figure}[!t]
   \centering
   \includegraphics[width=1.0\columnwidth]{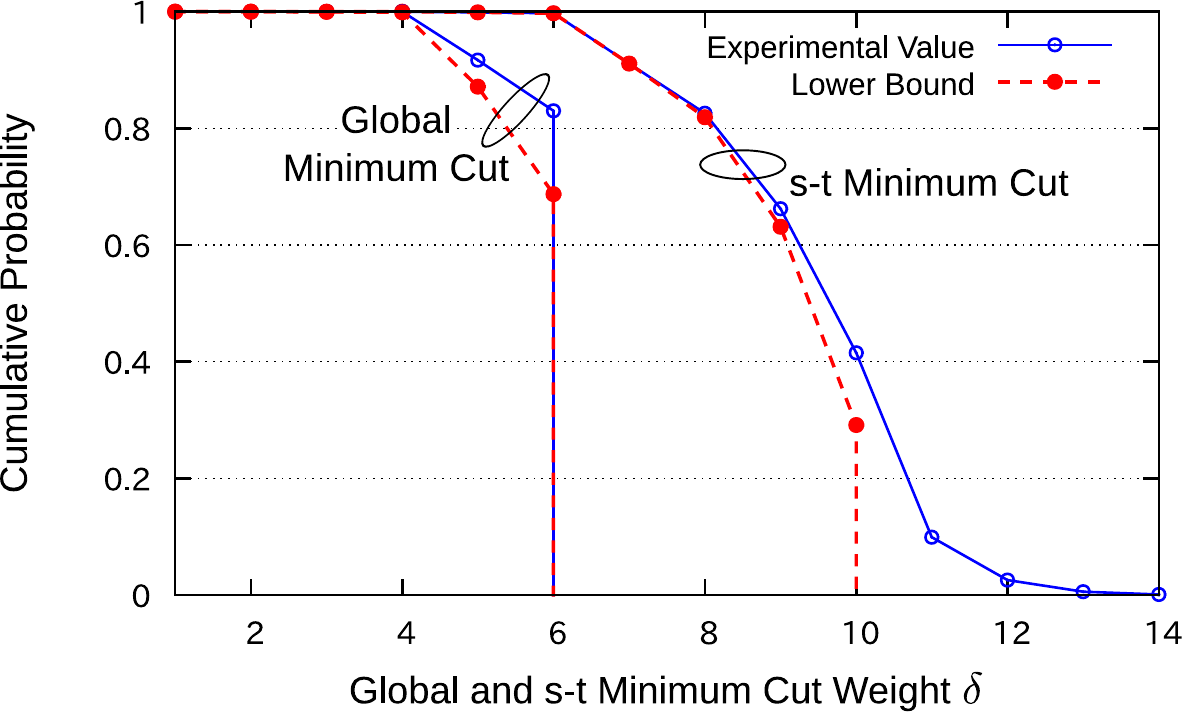}
   \caption{Accumulate distribution of the minimum $\st$ cut weight and
   the global minimum cut weight: experimental values and lower bounds.}
   \label{fig:experimental4}
  \end{figure}

  Figure \ref{fig:experimental1} presents the accumulate distribution of
  minimum $\st$ cut weight $\iProb{\lgst \ge \delta}$ of sparse and
  dense graph ensembles.  In the sparse case, the number of vertices and
  edges are $n = 120$ and $m = 248$. We assumed the degree distribution
  $\degree(x) = (1/3) x^3 + (1/3) x^4 + (1/5) x^5 + (2/15) x^6$. In the
  dense case, the parameters $n = 120, m = 488, \degree(x) = (1/3) x^7 +
  (1/3) x^8 + (1/5) x^9 + (2/15) x^{10}$ were assumed. The dashed lines
  represent values of the lower bound presented in Theorem
  \ref{theorem:maintheorem} and the solid lines present approximate
  values $\iProb{\lgst \ge \delta}$ obtained from computer
  experiments. From these experimental results, we can observe that the
  proposed lower bound captures the behaviors of the accumulate
  distribution $\iProb{\lgst \ge \delta}$ fairly well. Figure
  \ref{fig:experimental4} shows a comparison between the minimum $\st$
  cut and the global minimum cut weight. The lower bound for the global
  minimum cut weight is obtained according to the argument in
  \cite{Fujii-isita2012}.  In this case, the parameters $n=120$, $m =
  600$ and $\degree(x) = (1/24) x^6 + (1/24) x^7 + (1/12) x^8 + (1/6)
  x^9 + (1/3) x^{10} + (1/6) x^{11} + (1/12) x^{12}+ (1/24) x^{13} +
  (1/24) x^{14}$ were exploited.

  \section{Conclusion}

  In this paper, a lower bound on the accumulate distribution of the
  minimum $\st$ cut weight for a random graph ensemble is
  presented. From computer experiments, it is observed that the lower
  bound reflects actual statistical behavior of the minimum $\st$ cut
  weight. The proof technique used in this paper has close relationship
  to the analysis for average weight distribution of LDGM codes and it
  may be applicable to related problems on graphs such as the evaluation
  of the size of the minimum vertex cover over a random graph ensemble.

  % \section*{Acknowledgment}

  %% References:
  %% We recommend the usage of BibTeX:
  %%
  \nocite{}
  \bibliographystyle{IEEEtran}
  \bibliography{library,manual-library}
  %%
  %% where we here have assume the existence of the files
  %% definitions.bib and bibliofile.bib.
  %% BibTeX documentation can be obtained at:
  %% http://www.ctan.org/tex-archive/biblio/bibtex/contrib/doc/
  %%
  %%
  %%
  %% Or manual references (pay attention to consistency!):

\end{document}